\documentstyle[12pt]{article}
\newcommand{\beq}{\begin{equation}}
\newcommand{\eeq}{\end{equation}}
\begin{document}
\begin{titlepage}
\begin{flushleft}
       \hfill                      {\tt hep-th/9603081}\\
       \hfill                      USITP-96-03, UUITP-2/96\\
       \hfill                       March 1996\\
\end{flushleft}
\vspace*{3mm}
\begin{center}
{\LARGE D-particle Dynamics and Bound States\\ }
\vspace*{12mm}
{\large Ulf H. Danielsson\footnote{E-mail: ulf@teorfys.uu.se} \\
\vspace{4mm}
Gabriele Ferretti\footnote{E-mail: ferretti@teorfys.uu.se} \\
\vspace{4mm}
{\em Institutionen f\"{o}r teoretisk fysik \\
Box 803\\
S-751 08  Uppsala \\
Sweden \/}\\
\vspace*{5mm}
Bo Sundborg\footnote{E-mail: bo@physto.se} \\
\vspace{4mm}
        {\em Institute of Theoretical Physics \\
         Fysikum \\
        Box 6730\\
        S-113 85 Stockholm\\
        Sweden\/}\\}
\vspace*{10mm}
\end{center}

\begin{abstract}
We study the low energy effective theory describing the dynamics of
D-particles.
This corresponds to the quantum mechanical system obtained by dimensional
reduction of $9+1$ dimensional supersymmetric Yang-Mills theory to $0+1$
dimensions and can be interpreted as the non relativistic 
limit of the Born-Infeld
action. We study the system of two like-charged D-particles and find evidence
for
the existence of  non-BPS states whose mass grows like $\lambda^{1/3}$ over the
BPS mass. We give a string interpretation of this phenomenon in terms of a
linear
potential generated by strings stretching from the two D-particles. Some
comments
on the possible relations to black hole entropy and eleven dimensional
supergravity
are also given.
\end{abstract}

\end{titlepage}

\section{Introduction}

The last year has seen a tremendous activity in the subject of nonperturbative
string theory. Not only have fascinating dualities between various string
theories been proposed, but also we have obtained powerful tools to investigate
string solitons, collectively known as $p$-branes\footnote{$p$ is the dimension 
of the soliton: $p=0$ for particle-like objects, $p=1$ for strings etc.}, 
see e.g. \cite{DKL,D}. In
particular, those solitons arising in type IIA and IIB string theory and
carrying Ramond-Ramond (RR) charges can, according to \cite{P,PCJ}, be
described using  Dirichlet-branes (D-branes). The D-branes allow us to use
string theory tools to study objects that previously have been accessible only
in the field theory limit.

An exciting application is the study of black holes. Several low dimensional
examples have been
discussed in detail, see e.g. \cite{SV,GHKM,DM,CM,HS2,GKP}. By wrapping higher
dimensional $p$-branes around compact
cycles and considering bound states of such objects, black holes in the
compactified theory can be constructed. These studies have been particularly
successful in cases where the string coupling is constant over spacetime or at
least remains small near the horizon. This is the case for some dyonic black
holes with both magnetic and electric charges. Using the D-brane technology of
\cite{P,PCJ}
one can then reliably compute the number of states given their charges and
masses. As a
remarkable illustration of the power of string theory it is then found that the
corresponding entropy agrees with the prediction of Bekenstein-Hawking.

In this paper we will study D-particles ($0$-branes) in ten dimensional type
IIA string
theory. In e.g. \cite{HS} such solitonic objects have been constructed using
the
effective field theory equations. In particular there are extremal objects with
vanishing horizon area (and entropy) with mass $M=Q/\lambda$ in the string
metric\footnote{Throughout the paper we use dimensionless units by setting
$\alpha^\prime=1$.}. $Q$ is the  RR charge and $\lambda$ the asymptotic value of
the string
coupling. Since the theory also contains $6$-branes that are 
electromagnetic duals to
the D-particles, the analysis of Dirac implies that $Q$ must be quantized. In
two remarkable papers, \cite{T,W1} (see also \cite{HT,H}),
it has been argued that these objects also
have an eleven dimensional explanation. They correspond to Kaluza-Klein states
of eleven dimensional supergravity compactified on a circle with radius $R
\sim \lambda ^{2/3}$. Further results on higher dimensional D-branes in type IIA
can be found in \cite{BGPT,BRGPT}.

The extremal black holes have properties that identify them as BPS states. They
break half the supersymmetries and are also stable according to the classical
Bekenstein-Hawking analysis.
The BPS states are extremely important objects since it is possible to make
exact statements about their properties. However, from a physical point of
view, the non-BPS states are perhaps even more interesting. They are not
expected to be stable and given the Bekenstein-Hawking thermodynamic argument
they should decay through Hawking radiation. Excited, non-BPS D-particles were
briefly discussed in \cite{GHKM,KT}. For a single D-particle these states are
described by open strings with both ends attached to the D-particle. The only
possible excitations (in contrast to higher dimensional $p$-branes) are massive
modes of the string. For D-particles of higher charge, however, there are other
possibilities to construct non-BPS states. This is what  will be the main
subject of the paper.

The paper is organized as follows. In section two we consider the derivation of
effective theories for D-branes. The appearance of the Born-Infeld action is
discussed and it is argued that it is consistent to limit oneself to the study
of the Yang-Mills theory describing its non relativistic limit. In
section three we study in detail the supersymmetric
quantum mechanical problem describing a system of two bound
D-particles. We treat the problem in the Born-Oppenheimer approximation and
show the existence of non BPS states of mass $2/\lambda + \lambda^{1/3}
\epsilon$, where $2/\lambda$ is the BPS mass for $Q=2$ and $\epsilon$ the
eigenvalue of a one dimensional Schr\"odinger equation independent on
$\lambda$. In section four we present the string interpretation of the result.
We show how
each bound state can be interpreted as a particular configuration of straight 
open strings stretching between the D-particles. 
In section five we conclude with some
speculation about the possible role of eleven dimensional supergravity.

\section{$p$-branes from D-branes}

According to the D-brane prescription a $p$-brane with RR charges can be
described by an open string theory where the string end points are
restricted to lie on the D-brane. The effective field theory for this open
string theory living on the D-brane can be obtained by standard techniques, e.g
\cite{DLP,L}. The open strings have Neumann boundary conditions for $p$ of the
spatial coordinates and therefore couple to gauge fields within the D-brane.
For a single D-brane we simply get
a $U(1)$ gauge group.
What kind of gauge theory do we obtain? According to \cite{L}, 
see also \cite{T2,S}, we do not find the
familiar Yang-Mills theory but rather a Born-Infeld action. The bosonic part of
the effective action for a D-brane is
\beq
S_p= \int d^{p+1}\sigma e^{-\phi} {\rm Tr} \sqrt{ -{\rm det}( G+ F)}
\eeq
where $G$ is the induced metric on the D-brane and $F$ the field strength of
the $U(1)$ gauge field coupling to the open string.
For a D-particle this reduces to
\beq
S_0 = \frac{1}{\lambda} \int dt \sqrt{-{\rm det}G} = \frac{1}{\lambda} \int dt
\sqrt{1-v^2}
\eeq
for $e^{\phi} = \lambda$. The action is simply the action for a relativistic
particle with mass $M=1/\lambda$.

Another way to obtain this result is to start with the ten dimensional open
string
theory, i.e. a $9$-brane. In this case all string coordinates have Neumann
boundary conditions and couple to electromagnetic fields in the ten dimensional
space time. To obtain the action for a $p<9$-brane we must impose Dirichlet
boundary conditions \cite{G} in $9-p$ of the spatial directions. This can effectively be
taken care of by a dimensional reduction in the directions where we want
Dirichlet boundary conditions. We then need to reinterpret the corresponding
components of the gauge field as the new transversal coordinates. If, in the
ten dimensional action, we restrict ourselves to gauge potentials depending
only on time,
$A_{i} = A_{i} (t)$, the fully dimensionally reduced action is simply
$\frac{1}{\lambda} \int
dt \sqrt{1-\dot{A}^2}$. $A_i$ should now be thought of as the position of the
D-particle. From this point of view it becomes clear that the
Yang-Mills approximation to Born-Infeld is equivalent to considering
nonrelativistic D-particles. In fact, within the nonrelativistic approximation
we could as well have started with the ten dimensional Yang-Mills limit of the
Born-Infeld action. This is what we will do in the following section.

The parallel with special relativity was indeed the motivation for Born and
Infeld to introduce their action in the thirties \cite{BI}. As pointed out in
\cite{B},
the form of the Born-Infeld action is fixed through Neuman-Dirichlet duality.

For $n$ superimposed D-branes the gauge group to use, according to \cite{W2},
is $U(n)$. If we factor out
a $U(1)$ (it just corresponds to a common spatial translation of all the
D-particles) the relevant gauge group is instead $SU(n)$.
However, the non-abelian gauge bosons get masses if the superimposed D-branes
are pulled apart, that are proportional to the string tension and the distance
between the D-branes.  The physical picture of a W-boson is that of a string
stretching between the D-branes. We will find a nice illustration of this
phenomenon in the calculations of the following sections.
Eventually, when the D-particles are far enough apart, the theory becomes that
of
$n$ independent $U(1)$'s, one for each D-brane describing their positions in
space-time.

Presumably the full theory is some non-abelian supersymmetric generalization of
Born-Infeld dimensionally reduced to zero dimensions. We do not know how to
write
down such a theory. Instead we will follow \cite{W2}, see also \cite{S1,S2}, 
and start with $N=1$ SUSY
Yang-Mills in ten dimensions with gauge group $SU(n)$. In particular we will
study the case of $SU(2)$.
{}From our point of view, the use of Yang-Mills rather than Born-Infeld
corresponds  to a nonrelativistic approximation as explained
above\footnote{Strictly speaking, the identification of non-abelian Yang-Mills
theory as giving the non-relativistic approximation of the action for several
D-branes is threatened by higher order non-abelian terms \cite{T3,GW}. We
thank A. Tseytlin for bringing this fact to our attention. Such
terms do not affect our conclusions for small coupling $\lambda$.}. 
Further on, when we discuss the presence of bound states this 
means that we must restrict
ourselves to only light excitations as compared to the mass of the D-particles.

\section{The study of the effective theory}

In this section we derive and study the effective theory describing the
interaction
of two D-particles of equal charge. We derive the Schr\"odinger
equation by dimensional reduction and study it using the Born-Oppenheimer
approximation.
We calculate the spectrum and the degeneracies and find the existence of bound
states above the BPS ground state.

\subsection{Notation and conventions.}

We start by considering $N=1$ supersymmetric Yang-Mills theory in $9+1$
dimensional Minkowski
space with metric $ g_{\mu\nu} = {\rm diag}(+1, -1,\cdots, -1)$.
The field content is the gauge potential $A_\mu^a$ and a Majorana-Weyl spinor
$\Psi^a$ in the adjoint representation of the gauge group.
We fix the gauge group to be  $SU(2)$, and denote the structure constants by
$\epsilon^{abc}$. The $32\times 32$ dimensional Dirac matrices $\Gamma_\mu$
$\mu=0, \cdots, 9$, satisfying $\{\Gamma_\mu, \Gamma_\nu\} = 2g_{\mu\nu}$,
are written as
\begin{eqnarray}
       \Gamma_0 &\equiv&  \Gamma^0 = {\bf 1} \otimes \sigma_2 \nonumber\\
       \Gamma_i  &\equiv&  -\Gamma^i =  \gamma_i \otimes i\sigma_1\nonumber \\
       \Gamma_S &\equiv&  \Gamma^0\cdots\Gamma^9 = {\bf 1} \otimes
       \sigma_3,
\end{eqnarray}
where $\gamma_i$ are $9$ real symmetric $16\times 16$ dimensional matrices
satisfying $\{\gamma_i, \gamma_j\} = 2 \delta_{ij}$, ${\bf 1}$ is the
$16$ dimensional identity matrix and $\sigma_{1,2,3}$ the usual Pauli matrices.
The first $8$ matrices $\gamma$ can be identified with the Dirac matrices of
${\rm spin}(8)$ and the last with the $8$ dimensional chirality
$\gamma_9 = \gamma_1\cdots\gamma_8$.
The action is\footnote{In ordinary Yang-Mills theory $2\lambda=g^2$, with $g$
the Yang-Mills coupling. The factor of $2$ comes from considering the reduced
mass of the two D-particles, each of mass $1/\lambda$.}
\beq
       S={1\over{2\lambda}}\int d^{10} x\; \left( -{1 \over 4}
           F^{a\mu\nu} F^a_{\mu\nu} +
           {i \over 2}\bar\Psi^a \Gamma ^\mu D_\mu \Psi^a \right),
\label{actionten}
\eeq
where
\begin{eqnarray}
       F^a_{\mu\nu} &=& \partial_\mu A_\nu^a - \partial_\nu A_\mu^a +
       \epsilon^{abc} A^b_\mu A^c_\nu\nonumber \\
       D_\mu \Psi^a &=& \partial_\mu \Psi^a +  \epsilon^{abc} A_\mu^b \Psi^c
\end{eqnarray}
are the field strength and the covariant derivative respectively.
The supersymmetry transformation that leaves the action invariant can be
written
by introducing a constant anticommuting Majorana-Weyl spinor $\epsilon$
\begin{eqnarray}
        \delta A_\mu^a &=& i \bar\epsilon \Gamma_\mu\Psi^a\nonumber \\
        \delta\Psi^a &=& {1\over 2} F_{\mu\nu}^a \Gamma^{\mu\nu}\epsilon
        \label{susyten}
\end{eqnarray}
where $\Gamma^{\mu\nu} = [\Gamma^\mu, \Gamma^\nu]/2$ and the same notation will
be used below for the $\gamma_i$'s. The transformations
(\ref{susyten}) yield the supercurrent
\beq
       J^\rho = {i\over 2} \Gamma^{\mu\nu} \Gamma^\rho F_{\mu\nu} ^a\Psi^a.
\eeq

\subsection{Dimensional reduction}

The dimensional reduction \cite{BSS} of  (\ref{actionten}) all the way down
to $0+1$ dimensions
(supersymmetric quantum mechanics \cite{W3,W4,CKS}) can be easily
performed by letting all
the fields be independent of all space components $x_1,\cdots,x_9$. It is also
more convenient to work with the $16$ components spinors $\psi^a$ defined
through
\beq
        \Psi^a = \sqrt{2\lambda} \; \psi^a  \otimes \left (\matrix{ 1\cr 0\cr }\right ),
\eeq
rather than $\Psi^a$.
Representing the time derivative by a dot, the various components of the field
strength and the covariant derivative are
\begin{eqnarray}
        F_{0i}^a &=& \dot A^a_i + \epsilon^{abc} A_0^b A_i^c\nonumber \\
        F_{ij}^a &=& \epsilon^{abc} A_i^b A_j^c\nonumber \\
        D_0\psi^a &=& \dot\psi^a + \epsilon^{abc} A_0^b \psi^c\nonumber \\
        D_i\psi^a &=& \epsilon^{abc} A_i^b \psi^c.
\end{eqnarray}
Substituting in (\ref{actionten}) and dropping the volume term we obtain
\begin{eqnarray}
        S  &= & \int dt \; \Bigg[ {1\over{2\lambda}}
        \Bigg( {1\over 2} \dot A_i^{a2} + \epsilon^{abc}
         \dot A_i^a A_0^b A_i^c + {1\over 2} \left( \epsilon^{abc}A_0^b
               A_i^c\right)^2 - {1\over 4} \left( \epsilon^{abc} A_i^b
               A_j^c\right)^2 \Bigg) \nonumber \\
               &+ & {i\over 2} \psi^a\dot\psi^a + {i\over 2} \epsilon^{abc}
                \psi^a A_0^b\psi^c + {i\over 2} \epsilon^{abc}
                A_i^a\psi^b\gamma_i\psi^c \Bigg].
\end{eqnarray}
All space indices are now ``internal'' indices of this quantum system and will
always be written downstairs, with the convention that two repeated indices are
summed with the metric $+\delta_{ij}$. The transpose sign $^T$ on the spinor to
the left should also always be understood. As in any gauge theory, the $A_0$
component is an auxiliary field enforcing the Gauss law
\beq
        G^a(t) = {{\delta S }\over{\delta A_0^a(t)}} = 0.
\eeq
We shall work in the ``temporal gauge'' $A_0^a \equiv 0$ and denote by $E_i^a$
the momentum conjugate  to the $A_i^a$. We then quantize the theory by
introducing the canonical commutation and anti-commutation relations
\begin{eqnarray}
          [E_i^a, A_j^b] &=& -i \delta_{ij}\delta^{ab} \nonumber \\
          \{\psi^a_\alpha, \psi^b_\beta \} &=& \delta^{ab}\delta_{\alpha\beta}.
          \label{canonical}
\end{eqnarray}
In dimensionally reducing the action from $9+1$ to $0+1$ dimensions, we have
gone from $N=1$ to $N=16$ real supersymmetries, whose generators we denote by
$Q_\alpha$.

In the temporal gauge, the Gauss law $G^a$, the supercharges $Q_\alpha$ and the
Hamiltonian $H$ read
\beq
       G^a = \epsilon^{abc} A^b_i E^c_i - {i\over 2} \epsilon^{abc}
\psi^b\psi^c,
               \label{gaussgenerators}
\eeq
\beq
       Q_\alpha = \sqrt{2\lambda}
       \gamma_{i\alpha\beta}\psi^a_\beta E_i^a - {1\over {2\sqrt{2\lambda}}}
            \epsilon^{abc} \gamma_{ij\alpha\beta} \psi^c_\beta
            A_i^a A_j^b,  \label{supercharges}
\eeq
and
\beq
        H =  \lambda E_i^{a2} - {1\over 2} A_i^a K_i^a
                  +{1\over {8\lambda}} \left(\epsilon^{abc} A_i^b A_j^c\right)^2.
                    \label{hamiltonian}
\eeq
In the expression for the Hamiltonian we have lumped all the dependence on the
fermionic degrees of freedom into the operator
\beq
         K_i^a = i \epsilon^{abc} \psi^b\gamma_i\psi^c. \label{kappaia}
\eeq
The algebra satisfied by the above operators is
\begin{eqnarray}
        [G^a, G^b] &=& i\epsilon^{abc} G^c \nonumber \\
         {[}G^a, Q_\alpha] &=& 0 \nonumber \\
        \{ Q_\alpha, Q_\beta \} &=& 2\delta_{\alpha\beta} H -
              2\gamma_{i\alpha\beta} A_i^a G^a,
\end{eqnarray}
and thus the supersymmetry algebra is satisfied only weakly.

\subsection{General comments on the quantization procedure}

A physical way to study the system is to use the Born-Oppenheimer
approximation, treating the fast oscillations orthogonal to the classical
vacuum as the ``electrons'' and then restricting oneself on the minimum
of the bosonic potential (the ``nuclei''). Schematically, we shall write the
full wave function ${\bf \Psi}$ as
\beq
        {\bf \Psi} = {\bf \Xi}(\hbox{slow}) \otimes {\bf \Phi}(\hbox{fast,
        {\it slow}}), \label{bosplit}
\eeq
where ${\bf \Xi}$ is the wave function depending only on the coordinates of the
vacuum and  ${\bf \Phi}$ is the wave function along the non degenerate
directions, parametrized by the slow coordinates of the vacuum.
Unfortunately, as we shall see, the
approximation breaks down near the origin $A=0$ and this prevents one from
making a rigorous statement about the existence of the zero energy state. It
does however allow one to make some non trivial consistency checks and also,
perhaps more interestingly, to study the existence of excited states, where the
presence of a linearly binding potential at infinity cures the problems at the
origin by making it possible to find smooth and normalizable wave functions
within this approximation.

The reason why these excited states arise is that, in spite of the fact that
the system has flat directions, the potential away from these direction becomes
steeper and steeper as one moves to the large field region. In other words, a
classical particle rolling along the classical vacuum sees the walls of the
potential ``closing in''. This essentially prevents a wave packet, (or quantum
particle) from spreading to infinity. This effect is captured by the
Born-Oppenheimer approximation by showing that the quantization along the
non-flat directions gives rise to a linear potential that binds the particle.
This effect is of course not present for the ground state, where the bosonic
and fermionic contributions cancel out, the quantum mechanical analog of the
familiar field theoretical statement.

Let us also mention that, if the ground state exists and is unique, then, by
virtue of the commutation relations among (\ref{gaussgenerators}) and
(\ref{supercharges})
it is guaranteed to be gauge invariant.
This is so because if one defines the unitary operator
$U(\theta) = \exp(i\theta^aG^a)$ for $3$ arbitrary real numbers $\theta^a$
then, if
$Q_\alpha {\bf \Psi} = 0$, it is also
$Q_\alpha U(\theta) {\bf \Psi} \equiv U(\theta)  Q_\alpha {\bf \Psi} =0$. But,
since the ground state is unique, ${\bf \Psi} = U(\theta) {\bf \Psi}$ (up to an
irrelevant phase), i.e. ${\bf \Psi}$ gauge invariant. However, the gauge
invariance of the excited states is not given by this simple argument but it
will be imposed during quantization and it will pose some restrictions on the
degeneracy of the spectrum.

Usually, the presence of the first class constraints introduces another slight
complication; a gauge invariant state is, in general, not normalizable on the
full configuration space and normalizability should only be imposed after
having factored out
the gauge directions. Such problem does not arise in our case because, after
having fixed the temporal gauge, the only remnant of ``gauge" invariance is a
global $SU(2)$ transformation of finite volume.

\subsection{The classical ground state}

We begin our analysis with a more thorough investigation of the classical
ground state. We are dealing with a supersymmetric quantum mechanical system
described by $27$ bosonic coordinates $A_i^a$, $a=1,2,3$, $i=1,\cdots,9$
and $48$ fermionic ones $\psi^a_\alpha$, $a=1,2,3$,
$\alpha=1,\cdots,16$\footnote{Recall that supersymmetry is only realized on
shell, after eliminating $3$ of the bosonic coordinates.}.
It is easy to see that the bosonic potential
\beq
      V= {1\over {8\lambda}} \left(\epsilon^{abc} A_i^b A_j^c\right)^2
      \label{bosonicpotential}
\eeq
vanishes on what can be identified as an $11$ dimensional cone\footnote{In
order to avoid confusion, we should like to stress that this includes some
gauge equivalent configurations, it is not the ``moduli" space of the theory,
which is
simply ${\bf R}^9$.}. To see this, consider the mapping
\beq
          {\cal C} \equiv \left({\bf R}^9 \setminus \{{\bf 0}\} \right)\times
          {\bf S}^2\to {\bf R}^{27}, \label{almost}
\eeq
defined by
\beq
      A_i^a = \lambda_i n^a ,\quad \lambda_i\in {\bf R}, \quad n^a n^a =1
                    \quad r^2 \equiv \lambda_i^2 \not= 0.
                   \label{doublecover}
\eeq
Eq. (\ref{doublecover}) is obviously a minimum of (\ref{bosonicpotential}).
Moreover, the Hessian of (\ref{bosonicpotential}) at a generic point of
(\ref{doublecover}) is
\beq
        {{\partial^2 V}\over{\partial A_i^a \partial A_j^b}} = {1\over{2\lambda}}r^2
        \left(\delta_{ij} - {{\lambda_i \lambda_j }\over {r^2}}\right)
        \left(\delta^{ab} - n^a n^b \right),
\eeq
which has $16$ non degenerate directions and $11$ degenerate ones for
$r\not= 0$; (for $r=0$ all directions are degenerate and that represents the
apex
of the cone).

Hence, away from $r=0$ the map (\ref{almost}) is regular and describes
a two-to-one map of
$\left({\bf R}^9 \setminus \{{\bf 0}\} \right)\times S^2$
into ${\bf R}^{27}$. The reason why the map is two-to-one is because the points
$(\lambda_i, n^a)$ and $(-\lambda_i, -n^a)$ are mapped to the same point
of ${\bf R}^{27}$. One could mod out this ${\bf Z}_2$ factor from the sphere
but this is not necessary for our purposes, as long as we remember that
$\lambda_i \to - \lambda_i$ is a gauge transformation. Notice that the situation
generalizes to an arbitrary group $G$ in a straightforward way. If $d={\rm
dim}(G)$, $r={\rm rank}(G)$, then the bosonic vacuum is a $8r + d$  dimensional
cone, leaving a total of $8(d-r)$ non degenerate directions in the full space
of potentials ${\bf R}^{9d}$.

\subsection{Quantization along the non-degenerate directions.}

Let us first fix a point on  ${\cal C}$ (away from the apex) and
consider the quantization of the ``fast'' modes, i.e. the non degenerate
directions of the Hessian, in the same spirit as in the Born-Oppenheimer
approximation. All these points are in fact equivalent and, to fix the ideas,
we shall take the point on ${\cal C}$ described by $A_9^3=r$ and $A_i^a=0$
otherwise. By constructing the generic tangent vector to ${\cal C}$, it is easy
to see that the $16$ orthogonal directions are
$\delta A_i^a \equiv a_i^a$, non zero only for $a=1,2$ and $i=1, \cdots, 8$.
Denoting by $e_i^a$ their momenta, the Hamiltonian for the fast modes can be
written near that point as
\beq
      H_{\rm fast} = \sum_{a=1,2\;i=1\cdots 8} { (\lambda e_i^{a2} +
              {1\over{4\lambda}}r^2a_i^{a2})}
            - {r\over 2} K_9^3 \equiv H_B +H_F, \label{hfast}
\eeq
where $K_9^3$ is defined as in (\ref{kappaia}).

The first thing to notice is that the lowest eigenvalue of (\ref{hfast}) is
identically zero.
This comes from the fact that the zero point energy $16 \times (r/2) $ of the
$16$ bosonic oscillators in $H_B$  is cancelled by the lowest eigenvalue of the
fermionic part $H_F$. We will now study the spectrum in more details, taking
care of gauge invariance and will recover this fact as a simple consequence.
For now it only needs to be said that, away from the apex of ${\cal C}$ the
quantization of the fast modes yields the expected result: there is a unique
normalizable wave state ${\bf \Phi}$ such that  $ H_{{\rm fast}}{\bf \Phi} = 0$.

In order to study the spectrum in more details, we have to understand how gauge
invariance acts on the fast modes described by (\ref{hfast}). Two of the
generators of $SU(2)$ act on the coordinates of ${\cal C}$ and therefore should
not be considered here where the point on the vacuum has to be held fixed.  
They actually only determine how the fast modes at one point in the vacuum 
are related to fast modes at neighbouring points. But
the last generator, generically $ G=n^a G^a$, ($G = G^3$ in our case) acts in
the perpendicular direction to the vacuum and in the Born-Oppenheimer 
approximation we must project out all those
modes that do not satisfy $G{\bf \Phi}=0$, where $\bf \Phi$ is the wave 
function of the fast modes.

Let us begin with the bosonic part. The bosonic part $G_B$ of $G$ is the sum of
$8$ angular momentum operators, all commuting with (\ref{hfast})
\beq
        G_B = \sum_{j=1,\cdots 8} L^3_j, \qquad L^3_j = -i \epsilon^{3bc}
        a_j^b {{\partial}\over{\partial a_j^c}} \quad\hbox{(no sum over $j$ here)}.
\eeq
This means that the best way to think of the bosonic part of (\ref{hfast}) is
as
the sum of $8$ two dimensional harmonic oscillators $ H_B = \sum H_i$, each
characterized by two quantum numbers
$(N_i, m_i)$, $m_i = 0, \pm 2, \cdots \pm N_i$ if $N_i$ even,
$m_i = \pm 1, \pm 3, \cdots \pm N_i$ if $N_i$ odd. The complete bosonic part is
therefore labelled by $16$ quantum numbers
${\bf \Phi}_{N_1,m_1,\cdots,N_8,m_8}$, with total bosonic energy
$E_B = r(N_B + 8)$; $N_B=N_1 +\cdots +N_8$ and total angular momentum
$G_B = m_1 +\cdots +m_8$. It is then a straightforward combinatorial problem to
show that the degeneracy of a state with such total
energy and angular momentum is:
\beq
        d_B(N_B, G_B) = \left (\matrix{(N_B+G_B)/2 + 7\cr 7\cr}\right )
                  \left (\matrix{(N_B-G_B)/2 + 7\cr 7\cr}\right).
\eeq
(Note that $N_B\pm G_B$ is always an even integer.)

One might think that gauge invariance requires $G_B=0$ but one has also to take
into account the fermionic modes as well, and keep those states for which
$G=G_B + G_F=0$. The fermionic modes can be treated as follows: consider the
creation and annihilation operators
\begin{eqnarray}
       a_\alpha &=&\cases{
         {1\over{\sqrt{2}}}\Omega_{\alpha\beta}\left(\psi_\beta^1
                        + i \psi_\beta^2 \right) & for  $\alpha=1,\cdots 8 $\cr
         {1\over{\sqrt{2}}}\Omega_{\alpha\beta}\left(\psi_\beta^1
                        - i \psi_\beta^2 \right) & for $\alpha= 9, \cdots 16$
          \cr} \nonumber \\
       a_\alpha^\dagger &=& (a_\alpha)^\dagger
\end{eqnarray}
where $\Omega$ is an orthogonal matrix that diagonalizes
$\lambda_i\gamma_i$ ($=r\gamma_9$ in our case) as
$\Omega (\lambda_i\gamma_i)\Omega^T = r \hat\delta$, the diagonal matrix
$\hat\delta$ having matrix elements $\hat\delta_{\alpha\beta}
=+\delta_{\alpha\beta}$ for $\alpha\leq 8$ and $\hat\delta_{\alpha\beta}
=- \delta_{\alpha\beta}$ for $\alpha > 8$ .

In terms of $a$ and $a^\dagger$, the fermionic Hamiltonian and the fermionic
Gauss law are:
\beq
        H_F=r (a^\dagger_\alpha a_\alpha - 8) \equiv r(N_F -8); \quad
        G_F=- \hat\delta_{\alpha\beta} a_\alpha^\dagger a_\beta .
\eeq
The vacuum $|0>$ is manifestly gauge invariant and its zero energy cancels the
bosonic zero energy as promised. All other states can be constructed by acting
with $a^\dagger$'s on the vacuum. They can be labeled in terms of the two
quantum numbers $N_F=0,\cdots,16$ and $G_F=0,\pm 2,\cdots \pm N_F$ for
$N_F$ even and $G_F=\pm 1,\pm 3,\cdots \pm N_F$ for  $N_F$ odd. Their
degeneracy can also be easily computed as:
\beq
        d_F(N_F, G_F) = \left (\matrix{8 \cr (N_F+G_F)/2\cr}\right )
                      \left (\matrix{8 \cr (N_F-G_F)/2\cr}\right).
\eeq

It is now possible to combine these two results and to see that the gauge
invariant
fast modes are characterized by an overall energy $E=E_B+E_F=r N$, $N$ an even
integer and that the overall degeneracy of the gauge invariant sector is
\beq
        d(N) = \sum_{M=1,\cdots 16} \sum_{{\rm allowed} \;G^\prime}
        d_B(N-M, G^\prime) d_F(M,-G^\prime) \approx {{197\cdot199}
         \over{2^{14} \cdot 7!^2}} N^{14}.
\eeq
The first exact values are $d(0) = 1$, $d(2) = 192$, $d(4) = 11280$ and the
asymptotic formula is within $1 \%$ for $N>100$. The degeneracy is of course
only power law, as expected for a finite number of oscillators and should be
compared with the degeneracy of an unconstrained system of $16$ oscillators that
grows like $N^{15}$ instead of $N^{14}$. The energy $rN$ is the linear
potential
generated by the (unexcited) strings stretching between the two D-particles and
is responsible (for $N\not= 0$) for the existence of the excited bound states
to be discussed in the next section.

\subsection{Quantization along the flat directions}

The Hamiltonian $H_{\rm slow}$ on the vacuum is essentially the Laplacian on
the cone plus the linear effective potential found in the previous section. The
Laplacian can be computed by noticing that the induced metric on the
vacuum (\ref{doublecover}) is
\beq
        ds^2 = dA_i^{a2} = d\lambda_i^2 + \lambda_i^2 dn^{a2} =
         dr^2 + r^2 d\Omega_8 + r^2 d\Omega_2, \label{induced}
\eeq
where $d\Omega_8$ and $d\Omega_2$ are the metrics on the unit spheres.
Denoting by $L_8$ and $L_2$ the corresponding angular momenta, we obtain
\footnote{Recall that the reduced mass is $m_r = 1/2\lambda$.}:
\beq
        H_{\rm slow} = -{\lambda\over{r^{10}}}\partial_r r^{10} \partial_r +
          {\lambda\over{ r^2}}\left(L_8^2 + L_2^2\right) + Nr.
\eeq
This Hamiltonian can be simplified by recalling the familiar result that
the eigenvalues of the two angular momentum operators are given by the
quadratic Casimirs of $SO(3)$ and $SO(9)$ in the totally symmetric
representations, characterized by a single integer $l_2$ and $l_8$. Also, we
can rewrite the radial operator as
\beq
       -{1\over{r^{10}}}\partial_r r^{10} \partial_r =
       -{1\over{r^5}}\partial_r^2
         r^5 + {{20}\over{r^2}},
\eeq
and redefine the radial wave function as $y(r) = r^5 {\cal R}(r)$ to obtain the
one dimensional problem:
\beq
       -\lambda {{d^2}\over{dr^2}} y(r) +
        \left( \lambda{{20 + l_8(l_8 + 7) + l_2(l_2+1)}\over{r^2}}
          +N r \right) y(r) = E_{\rm tot} y(r). \label{onedim}
\eeq
Before continuing the calculation we would like to make three observations:

First, our approximation does not yield a bound state for $N=0$.
It is therefore impossible to draw any
rigorous conclusions about the zero energy state even though one could view the
cancelling of the energy for the fast modes and the good 
(square integrable at infinity)
asymptotic behavior of the Green function of the Laplacian as mild evidence in
favor of its existence.

A second point is that for $N\not =0$ the linear potential allows for the
existence of bound states within our approximation. Their energy scales like
$\lambda^{1/3}$. We interpret this as a small increase over the BPS
mass of the bound state ($2/\lambda$) valid at small coupling. These states are
therefore not BPS and most likely unstable in the full theory.

Finally notice that the ``effective angular momentum'' coming from the
reduction of the problem to the equivalent radial problem is $L_{\rm eff} = 20$
and not
$L_{\rm eff} = 12$ as one might guess by counting only the dimension of the
moduli space ${\bf R}^9$. In this context it should be noted that this 
possible effect is quantum mechanical in nature and therefore consistent 
with the vanishing 
classical force calculated in \cite{P} for D-particles at rest.

Now let us look more closely to the issue of gauge invariance of the wave
function for $H_{\rm slow}$. At first, it might seem that one should take $l_2
=0$, i.e., the wave function should be independent on $n^a$. In fact, due to
the presence of the fermions, it is possible to allow for an $n^a$ dependence
by considering the combination $n^a\psi^a$. As shown in the previous section,
at a generic point $(\lambda_i, n^a)$ one can use $32$ of the $48$ fermions to
make up the $16$ creation and annihilation operators needed for $H_{\rm fast}$.
This leaves $16$ fermions, generically $n^a\psi^a_\alpha$ that can be used to
make up the remaining $8$ creation and annihilation operators needed to match
the $24 = 27-3$ on shell bosonic degrees of freedom; they can be constructed as
\begin{eqnarray}
        a_{\hat \alpha} &=& {1\over{\sqrt{2}}} n^a(\psi^a_{\hat \alpha} +
          i \psi^a_{\hat \alpha +8}) \quad\hbox{for } \hat\alpha = 1,\cdots,
           8\nonumber \\
         a_{\hat \alpha}^\dagger &=& (a_{\hat \alpha})^\dagger.
\end{eqnarray}
We can act with up to $\tilde N_F= 8$ creation operators on the fermionic
vacuum. The result of such an operation will not be directly an eigenstate of
$L_2^2$ but it will contain all eigenstates $l_2\leq \tilde N_F$, $l_2$ even
(odd) if $\tilde N_F$ even (odd). However, each eigenstate can be easily
projected out by taking the traceless components of product $n^{a_1}\cdots
n^{a_{\tilde N_F}}$; e.g.
\beq
       n^a n^b \rightarrow (n^a n^b - (1/3) \delta^{ab}) + (1/3) \delta^{ab} =
       \{l_2 \equiv 2\} \oplus \{l_2 \equiv 0\} .
\eeq
By noticing that each allowed eigenvalue appears once in the reduction, we can
calculate the  degeneracies: For $l_2$ increasing from $0$ to $8$ we have:
$128$, $128$, $127$, $120$, $99$, $64$, $29$, $8$ and $1$. It is amusing to
note that the state with the lowest degeneracy is the state with the highest
value for $l_2$, an indication that the exact ground state
might have a large $l_2$ component.
In summary, the full wave function on ${\cal C}$ can be written  as
\beq
        {\bf \Xi} = {\cal R}(r) Y_{l_8, \vec{m}}(\Omega_8) {\cal P}_{l_2}
         \left( a^\dagger_{\hat\alpha_1}\cdots a^\dagger_{\hat\alpha_{\tilde
         N_F}} \right) |0>,
\eeq
where ${\cal P}_{l_2}$ is the projection described above, and $Y_{l_8,
\vec{m}}$
are the ``spherical harmonics'' on the eight-sphere\footnote{For completeness,
let us recall that the dimension of the first few such representations is $1$,
$9$, $44$, $156$, $450\cdots$ for $l_8=0$,  $1$, $2$, $3$, $4\cdots$.}.
By looking at (\ref{onedim}) we see that ${\cal R} \to 0$ as $r\to 0$ unless
$l_8=l_2=0$, so the wave function is continuous at the origin as it should.
Finally, in order for the wave function to be well defined after modding out
the ${\bf Z}_2$ symmetry we must have $l_2 + l_8 =$ even integer.

The one particle equivalent problem (\ref{onedim}) cannot be solved exactly
but, if the internal quantum numbers are such that the potential is not too
steep, its eigenvalues can be obtained through the WKB approximation in terms
of a (non degenerate) radial quantum number $n_r$. After scaling out the
dependence on $g$: $E_{\rm tot} = \lambda^{1/3} \epsilon$:
\beq
        \int_{x_{-}}^{x_{+}} \sqrt{(\epsilon - V(x)) } = \pi n_r
         \quad\hbox{where }\;
        V(x) = {{20 + l_8(l_8 +7) + l_2(l_2+1)}\over{ x^2}} + Nx, \label{epein}
\eeq
yielding, roughly, $\epsilon\approx (N n_r)^{2/3}$.

\section{String interpretation}

In treating the system of two D-branes as a dimensionally reduced $SU(2)$
Yang-Mills theory, solutions with gauge symmetry spontaneously broken to $U(1)$
describe branes which are separated from each other. Their separation can be
read off in the masses of the charged excitations, given by the ground state
energies of strings stretched between the two D-branes, and is proportional to
the distance between them. Since spontaneous symmetry breaking only works in
sufficiently high dimensions we expect quantum mechanical effects to modify the
picture for lower dimensional branes. Arguments based on duality also indicate
that D-particles in type IIA string theory should form (symmetric) bound states,
one for each RR charge. We have seen that there are additional bound states of
higher energies, at least in the Born-Oppenheimer approximation. We now wish to
test how this spectrum of excited states can be understood in the string
picture.

The Born-Oppenheimer approximation is adiabatic, which means that the slow
modes are treated as static on the time-scale of the fast modes. Only after the
effects of the fast modes has been taken care of does one study the motion of
the slow modes. Before the dynamics of the slow modes has been taken into
account, one essentially has the picture referred to above, with D-particles at
fixed positions. To the system of two D-particles can be added any number of
strings stretching between the two branes, or beginning and ending on the same
brane. (Some linear combinations of states with strings beginning and ending on
the same brane belong to the $U(1)$ describing the centre of mass motion of the
pair of D-particles, and not to the $SU(2)$ of the relative motion, which we
are focusing on here.)

In the low energy Yang-Mills approximation to the theory of open strings on the
D-particles only the string ground states are taken into account. In our case
they are simply the 8 vector components transverse to the straight strings
stretching between the two D-particles, in the bosonic sector, and the eight
$SO(8)$ spinor components in the fermionic sector. For each of these modes the
wave functions may have either positive or negative world-sheet parity, since
the strings are oriented. (The Chan-Paton factor may be symmetric or anti-symmetric.) 
Any number of such strings may be excited\footnote{For $n$ D-particles 
there are $8n(n-1)$ ways for bosonic straight strings to connect two D-particles, 
in agreement with the counting of  bosonic non-degenerate directions for 
$G=SU(n)$ at the end of section 3.4. }. We
recognize the spectrum of fast modes in section 3.5, by  interpreting $N_i$ as
the number of bosonic strings pointing in the $i$th transverse direction. The 
quantum number  $m_i$ counts the difference between the numbers 
of strings of positive and negative
parity. Since there are only 16 different fermion states (no massive
excitations in our approximation) there can be at most 16 fermionic strings
stretching between the D-particles. Of course, bound states of two D-particles
with an even number of fermionic strings will be bosons, but since the fermionic
strings do not form bosonic bound states by themselves, we get this bound
on the total number of fermionic strings stretching between the two branes.

Strings stretched between D-branes induce a linear potential between the 
branes, which causes them to move. The strength of the linear potential should 
be proportional to the number of strings connecting the D-particles. This is precisely 
what we see in the dynamics of the slow modes! We expect qualitatively similar 
effects in the dynamics of all Dirichlet $p$-branes with $p$ spatial dimensions curled 
up around a compact manifold. 

The Born-Oppenheimer approximation makes sense whenever the factorization in eq.
(\ref{bosplit}) results in an approximate additivity of energies from fast and slow modes, 
i.e. the fast wave function ${\bf \Phi}$ should vary much slower with respect to the slow 
parameters than with respect to the fast variables. This happens for 
$Nr \gg \lambda / r^2$. The self-consistency of the approximation depends on the
 shape of the approximate wave function. As noted before the fast modes must be
 in an excited state, and in addition the wave function of the slow modes must be 
concentrated at $r \gg \lambda^{1/3} $. This is most easily achieved by taking the 
angular momentum $l_8$ to satisfy $ l_8^2  \gg 1$.  

The bound states that we have found in the Born-Oppenheimer approximation have 
positive energies, and should be able to decay. Indeed, a four point coupling between 
two massive W particles and two photons in the Yang-Mills theory corresponds to an 
amplitude of order $\lambda$ between two strings stretching between fixed D-particles 
and two strings, one fixed to each of the two stationary D-particles. In the present 
Born-Oppenheimer framework part of these interactions are already included and 
generate the motion of the D-particles, but we suspect that one would also see 
decay processes where two stretched strings annihilate and their energy goes into 
kinetic energy of the D-particles, if one improves on the approximations.

\section{Conclusions}

We have shown, by studying the low energy effective action for two
D-particles, how it is possible to extract information about bound states above the
BPS threshold. These states have a rather simple interpretation as being
generated
by strings stretching between the D-particles. Since these states are not BPS
they
will most likely decay in the full theory and therefore should only be
interpreted as metastable. It should be possible to study their decay 
within the black hole picture. In this regard we
must keep in mind that the degeneracies obtained are only power law, as they
arise from a finite number of oscillators. They might contribute to subleading
corrections to the black hole entropy. 

It is also amusing to write down the
expression for the mass of the system using the
metric appropriate for eleven dimensional supergravity. This rescales the
masses by a factor $\lambda ^{1/3}$. If we further use the identification of
\cite{W1} where $R= \lambda ^{2/3}$ we find that
\beq
M= 2/R + R \epsilon
\eeq
where $R$ is the compactification radius in going from eleven dimensions to ten
and $\epsilon$ is the eigenvalue in
(\ref{epein}). (Note that this is only valid for $R$ small.) The simple
dependence on the compactification radius suggests that these states might have
a simple eleven dimensional explanation.

\section*{Note added}

After this paper was submitted a closely related paper by D. Kabat and 
P. Pouliot appeared, hep-th/9603127.

\section*{Acknowledgements.}

We wish to thank A. Alekseev, J. Kalkkinen, A. Niemi, B.
Nilsson, I. Pesando, H. Rubinstein and K. Sfetsos for helpful discussions.


\begin{thebibliography}{99}

 \bibitem{DKL}  M.J. Duff, R.R. Khuri and J.X. Lu,
    Phys. Rep. {\bf 259} (1995) 213, \hfill\break {\tt hep-th/9412184}.
\bibitem{D} M.J. Duff, Nucl. Phys. {\bf B442} (1995) 47, {\tt hep-th/9501030}.
\bibitem{P} J. Polchinski, Phys. Rev. Lett. {\bf 75} (1995) 4724,
       {\tt hep-th/9510017}.
\bibitem{PCJ} J. Polchinski, S. Chaudhuri and C.V. Johnson,
``Notes on D-branes", \hfill\break {\tt hep-th/9602052}.
\bibitem{SV} A. Strominger and C. Vafa, ``Microscopic Origin of the
Bekenstein-Hawking Entropy", {\tt hep-th/9601029}.
\bibitem{GHKM} S.S. Gubser, A. Hashimoto, I.R. Klebanov and J.M. Maldacena,
``Gravitational Lensing by $p$-branes", {\tt hep-th/9601057}.
\bibitem{DM} S.R. Das and S.D. Mathur, ``Excitations of D-strings, Entropy
and Duality", {\tt hep-th/9601152}.
\bibitem{CM} C.G. Callan and J.M. Maldacena, ``D-brane Approach to Black Hole
Quantum Mechanics", {\tt hep-th/9602043}.
\bibitem{HS2} G.T. Horowitz and A. Strominger, ``Counting States of
Near-Extremal Black Holes", {\tt hep-th/9602051}.
\bibitem{GKP} S.S. Gubser, I.R. Klebanov and A.W. Peet, ``Entropy and
Temperature of Black 3-branes", {\tt hep-th/9602135}.
\bibitem{HS} G.T. Horowitz and A. Strominger, Nucl. Phys. {\bf B360} (1991)
197.
\bibitem{T} P.K. Townsend, Phys. Lett. {\bf B350} (1995) 184, {\tt
hep-th/9501068}.
\bibitem{W1} E. Witten, Nucl. Phys. {\bf B443}, (1995) 85, {\tt
hep-th/9503124}.
\bibitem{HT} C.M. Hull and P.K. Townsend, Nucl. Phys. {\bf B438} (1995)
109, {\tt hep-th/9410167}.
\bibitem{H} C.M.  Hull, ``String Dynamics at Strong Coupling'', {\tt
hep-th/9512181}.
\bibitem{BGPT} E. Bergshoeff, M.B. Green, G. Papadopoulos and P.K.
Townsend, ``The IIA Super-eightbrane'', {\tt hep-th/9511079}.
\bibitem{BRGPT} E. Bergshoeff, M. de Roo, M.B. Green, G. Papadopoulos and P.K.
Townsend, ``Duality of Type II $7$-branes and $8$-branes'', {\tt hep-th/9601150}.
\bibitem{KT} I.R. Klebanov and L. Thorlacius, ``The Size of p-branes", 
{\tt hep-th/9510200}.
\bibitem{DLP} J. Dai, R.G.  Leigh and J. Polchinski, Mod. Phys. Lett. {\bf A4} 
 (1989) 2073.
\bibitem{L} R. G. Leigh, Mod. Phys. Lett. {\bf A4} (1989) 2767.
\bibitem{T2} P.K. Townsend, ``D-branes from M-branes", {\tt hep-th/9512062}.
\bibitem{S} C. Schmidhuber, ``D-brane Actions", {\tt hep-th/9601003}.
\bibitem{G} M.B. Green, Phys. Lett. {\bf B266} (1991) 325.
\bibitem{BI} M. Born and L. Infeld, Proc. R. Soc. {\bf 144} (1934) 425.
\bibitem{B} C. Bachas, ``D-brane Dynamics", {\tt hep-th/9511043}.
\bibitem{W2} E. Witten, ``Bound States of Strings and $p$-branes'', {\tt hep-th/9510135}.
\bibitem{S1}  A. Sen, ``A Note on Marginally Stable Bound States in Type II
String Theory'', {\tt hep-th/9510229}.
\bibitem{S2} A. Sen, ``U-Duality and Intersecting D-branes'', {\tt
hep-th/9511026}.
\bibitem{T3} A.A. Tseytlin, Nucl. Phys. {\bf B276} (1986) 391.
\bibitem{GW} D.J. Gross and E. Witten, Nucl. Phys. {\bf B277} (1986) 1.
\bibitem{BSS} L. Brink, J.H. Schwarz and J. Scherk,
       Nucl. Phys. {\bf B121} (1977) 77.
\bibitem{W3} E. Witten, Nucl. Phys. {\bf B188} (1981) 513.
\bibitem{W4} E. Witten, J. Diff. Geom. {\bf 17} (1982) 661.
\bibitem{CKS} F. Cooper, A. Khare and U. Sukhatme,
      Phys. Rep. {\bf 251} (1995) 267,\hfill\break
      {\tt hep-th/9405029}.

\end{thebibliography}
\end{document}